\documentclass[12pt,twoside]{article}
\usepackage{amsmath}

\newcommand{\bea}{\begin{eqnarray}}
\newcommand{\eea}{\end{eqnarray}}

\newcommand{\fract}[2]{{\textstyle\frac{#1}{#2}}}

\newcommand{\ham}{{\hat A_\mu}}
\newcommand{\han}{{\hat A_\nu}}

\newcommand{\pa}{\partial}

\newcommand{\pam}{\partial_\mu}
\newcommand{\pan}{\partial_\nu}
\newcommand{\paxa}{\partial x^\alpha}
\newcommand{\paxpm}{\partial x^{\prime\mu}}
\newcommand{\hAm}{\hat A_\mu}
\newcommand{\hApm}{\hat A'_\mu}
\newcommand{\mn}{{\mu\nu}}

\newcommand{\numeq}[2]{\begin{equation}
#2
\label{#1}
\end{equation}}
\newcommand{\refeq}[1]{(\ref{#1})}

\let\epsilon\varepsilon
\let\hat\widehat

\newcommand{\ct}{coordinate transformation}
\newcommand{\gt}{gauge transformation}
\newcommand{\nc}{noncommutative}
\newcommand{\nA}{non-Abelian}

\numberwithin{equation}{section}
\begin{document} 

\thispagestyle{empty}

\begin{flushright}
MIT-CTP-3208\\
BUHEP-01-29
\end{flushright}

\vspace{.5cm}

\begin{center}
{\LARGE \bfseries Covariant Coordinate Transformations\\[1ex]
 on Noncommutative Space}
\vspace{.8cm}  

R. Jackiw
\vspace{0.125cm}

\emph{Center for Theoretical Physics\\ 
Massachusetts Institute of Technology \\
Cambridge MA 02139-4307, USA}\\
\vspace{0.5cm}

S.-Y. Pi
\vspace{0.125cm}

\emph{Physics Department\\ 
Boston University \\
Boston MA 02215, USA}

\end{center}

\markboth{R. Jackiw and S.-Y. Pi}{Covariant Coordinate Transformations 
 on Noncommutative Space}
\vspace{.5cm}

\begin{abstract}\noindent
We show how to define gauge-covariant \ct s on a noncommuting space. The construction
uses the Seiberg-Witten equation and generalizes similar results for commuting coordinates.
\end{abstract}

%

\newpage 

\section*{Introduction}

Coordinate transformations in a gauge theory can be combined
with \gt s.  A ``gauge-covariant  \ct" is an especially interesting 
and useful combination of the two. Identified some years ago~\cite{1,2}, this
combination has now arisen in discussions of \nc\ gauge theories~\cite{3}. 

In our paper we investigate the action of \ct s on \nc\ vector gauge
potentials (connections) and tensor gauge fields (curvatures). We
establish that a \nc\ version of the gauge-covariant \ct\ is
particularly appropriate in this context. 

In the next Section, we review the relevant construction for
commutative \nA\ fields. Then the \nc\ story for U(1) fields is explained in
Section~2.

\section{Coordinate Transformations for\protect\newline Commuting Gauge Fields}
\label{CTGF}
Under a coordinate transformation
\numeq{e1.1}{
x^\mu \to x^{\prime\mu} = x^{\prime\mu} (x)
}
a covariant  vector field $A$ transforms as
\begin{gather}
A_\mu \to A'_\mu\nonumber\\
A'_\mu(x') = \frac{\pa x^\alpha}{\pa  x^{\prime\mu}} A_\alpha
(x)\ .
\label{e1.2}
\end{gather}
For infinitesimal transformations
\begin{gather}
x^{\prime\mu} (x) = x^\mu - f^\mu (x)\nonumber\\
\delta_f x^\mu \equiv  x^{\prime\mu} - x^\mu = -f^\mu(x)\ . 
\label{e1.3}
\end{gather}
\refeq{e1.2} implies
\numeq{e1.4}{
\delta_f A_\mu (x) \equiv
A'_\mu(x) -A_\mu(x) = f^\alpha(x) \frac\pa{\pa x^\alpha}
A_\mu(x) + \Bigl( \frac\pa{\pa x^\mu} f^\alpha(x)\Bigr)
A_\alpha(x) = L_f A_\mu
 }
where the last equality defines the Lie derivative, $L_f$, of a covariant
vector with respect to the (infinitesimal) transformation~$f$.

The composition law for the transformations is summarized by the
commutator algebra
\begin{align}
[\delta_f, \delta_g] x^\mu &= \delta_h x^\mu \label{e1.5}\\
[\delta_f, \delta_g] A_\mu &= \delta_h A_\mu \label{e1.6}
\end{align}
where $h$ is the Lie bracket of $f$ and~$g$:
\numeq{e1.7}{
h^\mu = g^\alpha \pa_\alpha f^\mu - f^\alpha \pa_\alpha g^\mu
\ .
 }

Taking $A$ to be a Hermitian \nA\ potential/connection
and defining the field strength/curvature $F$~by 
\numeq{e1.8}{
F_\mn = \pam A_\nu - \pan A_\mu - i[A_\mu, A_\nu]
} 
we find that for \emph{any} variation~$\delta$ of~$A$
\begin{align}
\delta F_\mn &= D_\mu \delta A_\nu -  D_\mu \delta A_\mu
\label{e1.9}\\
D_\mu  &= \pam - i[A_\mu, \quad ]\ .
\label{e1.10}
\end{align}
When the variation of $A$ is as in \refeq{e1.4}, $\delta_f
F$ becomes the appropriate Lie derivative
\numeq{e1.11}{
\delta_f F_\mn = f^\alpha \pa_\alpha F_\mn +
\pam f^\alpha F_{\alpha\nu}  + 
\pan f^\alpha F_{\mu\alpha} = L_f  F_\mn\ .
}
Also the algebra \refeq{e1.5}--\refeq{e1.7} is maintained:
\numeq{e1.12}{
[\delta_f, \delta_g] F_\mn = \delta_h F_\mn\ .
}

This of course is the entirely familiar geometrical story about
coordinate transformations. However, one may point out a defect:
gauge covariance is not maintained. This is especially evident in
\refeq{e1.11} where the gauge-covariant $F$ responds on the right
side with an ordinary derivative, which is not gauge covariant.  Further,
the response \refeq{e1.4} of the vector potential is noncovariant. 
This defect leads to  further awkwardness. For example, derivation of
conserved quantities associated with symmetries of dynamics by
Noether's theorem produces gauge-noninvariant entities that must
be ``improved''~\cite{4}.  Also, recognizing symmetric connections and
curvatures, i.e., configurations that are invariant under specific
\ct s, is obscured by the gauge  dependence~\cite{5}.

The gauge-covariant \ct\ is introduced to overcome the
above-mentioned defect. Observe that \refeq{e1.4} may be
identically presented as 
\begin{align}
\delta_f A_\mu &= f^\alpha\bigl(\pa_\alpha A_\mu - \pam
A_\alpha - i[A_\alpha,A_\mu] \bigr)
+ f^\alpha \pam A_\alpha + i f^\alpha
[A_\alpha,A_\mu] + 
\pam f^\alpha A_\alpha \nonumber\\
&= f^\alpha F_{\alpha\mu} + D_\mu (f^\alpha A_\alpha)\ .
\label{e1.13}
\end{align}
The last element is recognized as a \gt\footnote{Formula \protect\refeq{e1.13} in the
Abelian case is an instance of the well-known identity $L_f = i_f d + d i_f$, where $i_f$ is
the evaluation on the vector~$f$.}. Thus if we supplement the diffeomorphism generated
by~$f$ with a
\gt\  generated by
$\Lambda_f = - f^\alpha A_\alpha$ (a freedom on which we can
rely in a gauge theory)  we arrive at the gauge-covariant
transformation
\numeq{e1.14}{
\bar\delta_f A_\mu = f^\alpha F_{\alpha\mu}\ .
}
Moreover, when the above gauge-covariant response of $A$ is used in
\refeq{e1.9}, a gauge-covariant response is experienced by~$F$:
\numeq{e1.15}{
\bar\delta_f F_\mn = f^\alpha D_\alpha F_\mn + \pam f^\alpha
F_{\alpha\nu} + \pan f^\alpha F_{\mu\alpha} 
}
which again differs from the conventional Lie derivative expression
\refeq{e1.11} by a \gt\ generated by $\Lambda_f$. We call
\refeq{e1.15} the gauge-covariant Lie derivative.

Adoption of the transformation  rules \refeq{e1.14} and
\refeq{e1.15} simplifies the task of finding gauge-invariant
conserved quantities~\cite{4}, of recognizing symmetric configuration~\cite{5}, and
other advantages ensue as well~\cite{3}. However, the composition algebra
is modified. One finds
\begin{align}
[\bar\delta_f, \bar\delta_g]A_\mu & =
 \bar\delta_h A_\mu + D_\mu (g^\alpha f^\beta
F_{\alpha\beta})\label{e1.16}\\
[\bar\delta_f, \bar\delta_g] F_\mn & =
 \bar\delta_h F_\mn +  i [g^\alpha f^\beta
F_{\alpha\beta},F_\mn]\ .\label{e1.17}
\end{align}
In both instances the algebra closes up to a gauge transformation
generated by $g^\alpha f^\beta F_{\alpha\beta}$. Thus the 
gauge-covariant \ct\ rules provide a representation up to \gt s
of the \ct s~\cite{6}.

\section{Coordinate Transformations for\protect\newline Noncommuting Gauge Fields}

When coordinates do not commute, the formulas in the previous
Section present questions about how factors should be ordered. We
shall answer such questions by adopting simplifying
\emph{Ans\"atze}, and by some analysis.

To begin, we presume that the relation \refeq{e1.1} between
$x'$ and $x$ is at most linear; i.e., it is affine. Then
$\paxa/\paxpm$ is constant and Eq.~\refeq{e1.2} may be taken
over for a noncommuting vector field~$\hAm$
\numeq{e2.1}{
\hApm (x') = \frac{\paxa}{\paxpm} \hat A_\alpha (x)\ .
}
(Below we shall find a second reason for restricting the the
transformations to be affine.)

Next we wish to derive the \nc\ analog to \refeq{e1.4}.  We define
the ordering of~$x$ within $\hAm(x)$ by the Weyl procedure
\numeq{e2.2}{
\hAm(x) = \int_p e^{-ip\cdot x} a_\mu (p)\ .
}
Here $a_\mu (p)$ is a classical, $c$-number function and the
ordering of the \nc\ $x$ is defined by expanding the exponential.
It follows that infinitesimally
\begin{align}
\hApm (x') &\approx \hApm (x-f) =
                           \int_p e^{-ip(x-f)} a'_\mu (p)\nonumber\\
 &= \int_p e^{i\frac{p\cdot f}2} e^{-ip\cdot x} e^{i\frac{p\cdot
f}2}  a'_\mu (p)\nonumber\\
&\approx 
\int_p \Bigl(1+\frac i2 p\cdot f\Bigr) e^{-ip\cdot x}
\Bigl(1+\frac i2 p\cdot f\Bigr)  a'_\mu (p)\nonumber\\[1ex]
&\approx \hApm(x) + \fract12 f^\alpha \pa_\alpha \ham  + 
\fract12  \pa_\alpha \ham  f^\alpha\ .
\label{e2.3}
\end{align}
That the second integral equals the first follows from the
Baker-Hausdorff formula and the linearity of~$f$ in~$x$.

As a consequence of the above, we adopt \refeq{e1.4} to the \nc\
case as 
\numeq{e2.4}{
\delta_f \hAm = \fract12 \bigl\{ f^\alpha, \pa_\alpha \hAm
\bigr\}_\star + \pam f^\alpha \hat A_\alpha \ .
 }
Henceforth,  we view all quantities as commuting $c$-numbers, but
all products are ``star'' products, defined by 
\numeq{e2.5}{
(O_1 \star O_2) (x) = e^{\frac i2 \theta^{\alpha\beta}
\pa_\alpha^{\mathstrut} \pa_\beta'}  O_1(x) O_2(x')\bigr|_{x'\to
x}\ .
}
In \refeq{e2.4}, the curly bracket is the ``star'' anticommutator;
later we shall also use a square bracket to denote the
``star'' commutator: $[\ \,,\ ]_\star$. In the last entry  of \refeq{e2.4}, $\pam f^\alpha$ is
$x$-independent, hence no star product is needed. Also we restrict
the discussion to \nc\ U(1) gauge theories, where the potential
responds to a \gt\ as
\begin{align}
\ham \to \ham^\lambda &= (e^{i\lambda}) \star (\ham + i\pam) \star 
(e^{i\lambda})^{-1}\nonumber\\
 &\approx \ham + \pam \lambda - i[\ham,\lambda]_\star = \ham + D_\mu \lambda
\label{e2.6}\\[1ex]
D_\mu &= \pam - i[\ham, \ ]_\star
\label{e2.6.7}
\end{align}
and the field strength is constructed so that its response is covariant:
\numeq{e2.7}{
\hat F_{\mu\nu} = \pam \hat A_\nu - \pan \hat A_\mu - i [\hat A_\mu,\hat A_\nu]_\star
}
\begin{align}
\hat F_{\mu\nu}\to \ 
&(e^{\lambda}) \star \hat F_{\mu\nu} \star 
(e^{i\lambda})^{-1}\nonumber\\
 &\approx \hat  F_\mn - i [\hat F_\mn, \lambda]_\star \ .
\label{e2.8}
\end{align}

Formula \refeq{e2.4} suffers from  noncovariance defects in two ways. First, as in the
commutative case, the response involves the vector potential and is not gauge covariant.
Second, multiplication by~$x$ (which is present in $f^\alpha$) is not covariant, in the sense
that if $\Phi$ is a gauge-covariant quantity, $x^\mu\star \Phi$ is not. In order that covariance
be preserved $x^\mu$ should be supplemented by
$\theta^\mn \han$; that is, rather than multiplying by $f^\alpha$ one should 
multiply by
\numeq{e2.9}{
\hat f^\alpha = f^\alpha + \pa_\beta f^\alpha \theta^{\beta\gamma} \hat A_\gamma\ .
}
There does not seem to be a way to remedy these two defects if one remains
with~\refeq{e2.1}. But now we adopt the Seiberg-Witten viewpoint~\cite{7}, and discover a
way of constructing a covariant transformation law, which generalizes \refeq{e1.14} to the
\nc\ situation.

In the Seiberg-Witten framework, we view $\hat A$ to be a function of~$\theta$ and of the
commutative~$A$ (and its derivatives). This comes about because the $\theta$-dependence
of $\ham$ is governed by the Seiberg-Witten equation
\numeq{e2.10}{
\frac{\pa\ham}{\pa \theta^{\alpha\beta}} = 
-\fract18 \bigl\{  
\hat A_\alpha, \pa_\beta \ham + \hat F_{\beta\mu}
\bigr\}_\star -(\alpha \leftrightarrow \beta)\ .
}
Since the equation is of the first order, a solution is determined by specifying an initial
condition for $\hat A$  at $\theta=0$, which is  just~$A$. Although we do not need to
know the explicit solution to \refeq{e2.10}, for definiteness we record it to lowest order
in~$\theta$:
\numeq{e2.11}{
\ham = A_\mu - \fract12 \theta^{\alpha\beta} 
  A_\alpha  \bigl(\pa_\beta A_\mu + F_{\beta\mu}
\bigr) + \cdots\ .
}

We now understand that if we accept that $\hat A$ transforms under  affine
transformations as a covariant vector, so that \refeq{e2.1} and \refeq{e2.4}
are true, then in the Seiberg-Witten expression for $\hat A$ one must transform $A$
as a covariant vector and $\theta$ as a contravariant tensor; infinitesimally
\numeq{e2.12}{
\delta_f \theta^{\alpha\beta} = f^\gamma \pa_\gamma \theta^{\alpha\beta} 
- \pa_\gamma f^\alpha \theta^{\gamma\beta}
-\pa_\gamma f^\beta \theta^{\alpha\gamma}\ .
}
[This is seen explicitly in the approximate formula  \refeq{e2.11}.] Moreover,
since~$\theta$ is constant and should remain constant, the first term on the right
in \refeq{e2.12} vanishes, and the remaining terms will not generate an
$x$-dependence, provided~$f$ is affine in~$x$. (This gives the second reason for
restricting transformations to be affine.)

We now assert the principle that $\theta$ should not be transformed, because it is not a
dynamical variable; rather  it functions as a background. Thus to get the proper infinitesimal 
transformation for $\hat A$, we must subtract from \refeq{e2.4} the infinitesimal transformation
of~$\theta$~\refeq{e2.12}. We assert therefore that the desired infinitesimal \ct\ for $\hat
A$ is not
\refeq{e2.4}, but rather
\begin{subequations}\label{e2.13}
\numeq{e2.13a}{
\Delta_f\ham = \fract12 \bigl\{f^\alpha, \pa_\alpha \ham\bigr\}_\star 
+ \pam f^\alpha \hat A_\alpha - \frac{\pa\ham}{\pa\theta^{\alpha\beta}} \delta_f
\theta^{\alpha\beta}
 }
which reads, once \refeq{e2.10} and \refeq{e2.12} have been used, 
\numeq{e2.13b}{
\Delta_f\ham = \fract12 \bigl\{f^\alpha, \pa_\alpha \ham\bigr\}_\star 
+ \pam f^\alpha \hat A_\alpha 
- \fract14 \bigl\{\hat A_\alpha, \pa_\beta\ham + F_{\beta\mu}\bigr\}_\star
(\pa_\gamma f^\alpha \theta^{\gamma\beta} + \pa_\gamma f^\beta
\theta^{\alpha\gamma})\ .
 }
\end{subequations}
Note that just as $\delta_f \ham$ in \refeq{e2.4} this suffers from  the same
noncovariance defects. 

A straightforward manipulation shows that the above may also be written as
\numeq{e2.14}{
\Delta_f\ham = \fract12 \bigl\{\hat f^\alpha, \hat F_{\alpha\mu}\bigr\}_\star 
+ D_\mu \fract12 \bigl\{ f^\alpha + \fract12  \pa_\omega f^\alpha \theta ^{\omega\phi} \hat
A_\phi, \hat A_\alpha\bigr\}_\star\ .
 }
This shows that when a \ct\ is effected on $\hat A$ by only transforming $A$ as a vector
in the Seiberg-Witten expression for $\hat A$, and not transforming $\theta$, then the
response is the covariant first term on the right side of
\refeq{e2.14} and a \gt. 
As in the commutative case, we can supplement the coordinate transformation $\Delta_f$
with a \gt\ generated by 
$ \hat \Lambda_f   = - \fract12 \bigl\{ f^\alpha + \fract12  \pa_\omega f^\alpha \theta
^{\omega\phi} \hat A_\phi, \hat A_\alpha\bigr\}_\star$ and define
 the gauge-covariant \ct\ $\hat \delta_f$ on  a \nc\
covariant vector by
\numeq{e2.15}{
\hat\delta_f \ham = \fract12 \bigl\{\hat f^\alpha , \hat F_{\alpha\mu} \bigr\}_\star\ .
}
This is the desired generalization of \refeq{e1.13}. It is very satisfying that both defects
have been removed in the response \refeq{e2.15} -- the right side is covariant since it
involves multiplication by~$\hat f^\alpha$ (not $f^\alpha$) and there occurs the field
strength/curvature $\hat F$ (not the potential/connection~$\hat A$). 

Next we look to the transformation law for $\hat F$. Formula \refeq{e1.9} still holds in
the \nc\ framework because we do not vary $\theta$ in the $\star$-commutator
contributing to 
$\hat F$. Using \refeq{e2.15} for the variation of $\hat A$, we arrive at
\begin{align}
\hat\delta_f\hat F_\mn &= \fract12 \bigl\{\hat f^\alpha, D_\alpha \hat F_\mn\bigr\}_\star 
+ \pam f^\alpha \hat F_{\alpha\nu} 
+ \pan f^\alpha \hat F_{\mu\alpha}\nonumber\\ 
&\qquad{}+ \fract12 \pa_\alpha f^\gamma \theta^{\alpha\beta}
\Bigl( \bigl\{ \hat F_{\mu\beta}, \hat F_{\gamma\nu}\bigr\}_\star
- \bigl\{ \hat F_{\nu\beta}, \hat F_{\gamma\mu}\bigr\}_\star \Bigr)
\ .
\label{e2.16}
\end{align}
The first three terms on the right provide a natural \nc\ generalization of the gauge-covariant
Lie derivative of \refeq{e1.15}. But the last term gives an addition. This addition is analyzed
further by  defining
\begin{subequations}\label{e2.17}
\numeq{e2.17a}{
\theta_f^{\gamma\beta} = \pa_\alpha f^\gamma \theta^{\alpha\beta}
}
and recognizing that the part antisymmetric in $(\gamma,\beta)$ is proportional to the
variation
\refeq{e2.12} of~$\theta$. Therefore
\numeq{e2.17b}{
\theta_f^{\gamma\beta} = -\fract12 \delta_f \theta^{\gamma\beta} +
\theta_f^{(\gamma\beta)} 
}
\end{subequations}
with $\theta_f^{(\gamma\beta)} $ being the symmetric part. Substituting  \refeq{e2.17b}
into \refeq{e2.16} and noting that last term in parentheses is antisymmetric in
$(\gamma,\beta)$, we are left with 
\numeq{e2.18}{
\hat\delta_f\hat F_\mn = \fract12 \bigl\{\hat f^\alpha, D_\alpha \hat F_\mn\bigr\}_\star 
+ \pam f^\alpha \hat F_{\alpha\nu} 
+ \pan f^\alpha \hat F_{\mu\alpha} 
- \fract12 \delta_f \theta^{\alpha\beta}
 \bigl\{ \hat F_{\alpha\mu}, \hat F_{\beta\nu},\bigr\}_\star
\ .
 }
(We emphasize that no transformation of $\theta$ is carried out. Eq.~\refeq{e2.18} arises
because a certain  combination of~$f$ and~$\theta$ combines into an expression identical
with $\delta_f \theta^{\alpha\beta}$.)

Thus we conclude that those affine \ct s generated by~$f$ that also leave~$\theta$ invariant
($\delta_f
\theta =0$) transform the curvature by a  gauge-invariant, \nc\ Lie derivative
\numeq{e2.19}{
\hat\delta_f \hat F_\mn = \fract12 \bigl\{\hat f^\alpha, D_\alpha \hat
F_\mn\bigr\}_\star  + \pam f^\alpha \hat F_{\alpha\nu} 
+ \pan f^\alpha \hat F_{\mu\alpha} \qquad (\delta_f \theta
=0)
\ .
}

Now consider the algebra of these transformations. From \refeq{e2.15} and \refeq{e2.16}
we compute
\begin{multline}\label{e2.20}
[\hat\delta_f, \hat \delta_g]\ham = \hat\delta_h \ham 
+ D_\mu \Bigl(
\fract18 \bigl\{\hat g^\alpha, \{\hat f^\beta,\hat F_{\alpha\beta}\}_\star\bigr\}_\star
+\fract18 \bigl\{\hat f^\beta, \{\hat g^\alpha,\hat F_{\alpha\beta}\}_\star\bigr\}_\star
\Bigr)\\
{+\frac i8 \pa_\alpha f^\omega \theta^{\alpha\beta} \pa_\gamma g^\phi
\theta^{\gamma\delta} \Bigl(\bigl[\hat F_{\omega\phi}, D_\beta \hat F_{\mu \delta} +
D_\delta
\hat F_{\mu\beta} \bigr]_\star} 
+ \bigl[\hat F_{\beta\delta}, D_\omega \hat F_{\mu \phi} +
D_\phi
\hat F_{\mu\omega} \bigr]\\
+2 \bigl[\hat F_{\omega\delta}, D_\beta \hat F_{\phi\mu}\bigr]
-2 \bigl[\hat F_{\phi\beta}, D_\delta \hat F_{\omega\mu}\bigr]
\Bigr)\ .
\end{multline}
In the general case the algebra does not close, not even up to a \gt: The first term on the
right, where 
$ \hat h^\alpha = h^\alpha  + \pa_\beta h^\alpha \theta^{\beta\gamma} A_\gamma$,
with~$h$ given by the Lie bracket \refeq{e1.7} is needed for closure; the second, \gt\ term is
the \nc\ generalization of the \gt\ in \refeq{e1.16}; the last term spoils closure. However,
when the expressions involving $\theta$  are written with the help of \refeq{e2.17},
\refeq{e2.20} becomes
\begin{multline}\label{e2.21}
[\hat\delta_f, \hat \delta_g]\ham = \hat\delta_h \ham  
+ D_\mu \biggl(
\fract18 \bigl\{\hat g^\alpha, \{\hat f^\beta,\hat F_{\alpha\beta}\}_\star\bigr\}_\star
+\fract18 \bigl\{\hat f^\beta, \{\hat g^\alpha,\hat F_{\alpha\beta}\}_\star\bigr\}_\star
\\
 -\frac i{16} \Bigl(
\delta_f\theta^{\alpha\beta} \theta_g^{(\gamma\delta)} + \theta_f^{(\alpha\beta)}
\delta_g\theta^{\gamma\delta} 
- \fract12 \delta_f\theta^{\alpha\beta} \delta_g \theta^{\gamma\delta} \Bigr) 
[\hat F_{\alpha\delta}, \hat F_{\gamma\beta}]_\star \biggr)
\\
-\frac i8 
\delta_f\theta^{\alpha\gamma} \delta_g \theta^{\beta\delta} 
 \bigl[\hat F_{\alpha\beta},  D_\gamma \hat F_{\delta\mu} + 
      D_\delta \hat F_{\gamma\mu} \bigr]_\star
\ .
\end{multline}
Thus if again we restrict the coordinate transformations to those that leave $\theta$ invariant, 
$\delta_{f,g} \theta = 0$, the algebra closes up to \gt s, in complete analogy to~\refeq{e1.16}:
\begin{multline}\label{e2.22}
[\hat\delta_f, \hat \delta_g]\ham = \hat\delta_h \ham  
+ D_\mu \Bigl(
\fract18 \bigl\{\hat g^\alpha, \{\hat f^\beta,\hat F_{\alpha\beta}\}_\star\bigr\}_\star
+\fract18 \bigl\{\hat f^\beta, \{\hat g^\alpha,\hat F_{\alpha\beta}\}_\star\bigr\}_\star
\Bigr)\\
 (\delta_{f,g} \theta = 0)\ .
\end{multline}

The commutator acting on $\hat F_\mn$ behaves similarly. We record only the formula that
holds when $\theta$ is left invariant, that is, when \refeq{e2.19} is true:
\begin{multline}\label{e2.23}
[\hat\delta_f, \hat \delta_g]\hat F_\mn = \hat\delta_h \hat F_\mn  
+ \frac i8  \Bigl[  \bigl\{\hat g^\alpha, \{\hat f^\beta,\hat
F_{\alpha\beta}\}_\star\bigr\}_\star + \bigl\{\hat f^\beta, \{\hat g^\alpha,\hat
F_{\alpha\beta}\}_\star\bigr\}_\star,  \hat F_\mn
\Bigr]_\star \\
 (\delta_{f,g} \theta = 0)\ .
\end{multline}
Evidently this is the \nc\ generalization of \refeq{e1.17}.

Finally, we  examine how our transformations look in the context of the Seiberg-Witten
map. We begin with the $O(\theta)$ solution in \refeq{e2.11} and its consequence
\numeq{e2.24}{
\hat F_\mn 
= F_{\mu\nu}
+ \theta^{\alpha\beta} F_{\alpha\mu} F_{\beta\nu}
-\theta^{\alpha\beta} A_\alpha \pa_\beta F_{\mu\nu} + \cdots \ .
}
We transform the left side of \refeq{e2.11} according to \refeq{e2.15} and use \refeq{e2.24}
to express everything in terms of commuting variables to $O(\theta)$. In this way we get 
\begin{subequations}\label{e2.25}
\numeq{e2.25a}{
\hat\delta_f \ham = f^\gamma F_{\gamma\mu} +  f^\gamma \theta^{\alpha\beta} 
(F_{\alpha\gamma} F_{\beta\mu} + A_\beta\pa_\alpha F_{\gamma\mu} )
 + \pa_\alpha f^\gamma \theta^{\alpha\beta} A_\beta F_{\gamma\mu} + \cdots \ .
}
Correspondingly, the right side of \refeq{e2.11} is transformed according to \refeq{e1.14}
and \refeq{e1.15}:
\begin{multline}\label{e2.25b} 
\bar\delta_f \bigl(A_\mu - \fract12 \theta^{\alpha\beta} A_\alpha (\pa_\beta A_\mu +
F_{\beta\mu}) \bigr)  =\\
 f^\gamma F_{\gamma\mu}  +  \fract12 f^\gamma \theta^{\alpha\beta} 
\bigl(F_{\alpha\gamma} (\pa_\beta A_\mu +
F_{\beta\mu}) +  A_\beta \pa_\alpha F_{\gamma\mu} +
 A_\beta \pa_\gamma F_{\alpha\mu} \bigr)\\
 + \theta^{\alpha\beta} A_\beta (\pa_\alpha f^\gamma F_{\gamma\mu} + \fract12 \pam
f^\gamma F_{\alpha\gamma})+ \cdots \ .
\end{multline} 
\end{subequations}
Forming the difference between \refeq{e2.25a} and \refeq{e2.25b} we find that it
equals\linebreak 
$\pam (\fract12 f^\gamma \theta^{\alpha\beta} F_{\alpha\gamma} A_\beta)$; that is, the
two ways of effecting the coordinate gauge-covariant transformation coincide up to a \gt.

A similar result emerges when transformations of the field strength are compared. We begin
with \refeq{e2.24} and transform the left side according to \refeq{e2.19} and express
everything in terms of commuting variables with the help of \refeq{e2.11} and \refeq{e2.24}.
One finds 
\begin{subequations}\label{e2.26}
\begin{multline}\label{e2.26a}
\hat\delta_f \hat F_\mn = f^\gamma \pa_\gamma\hat F_\mn + \pam f^\gamma
\hat F_{\gamma\nu} + \pan f^\gamma \hat F_{\mu\gamma}\\
+ f^\gamma \theta^{\alpha\beta} 
\pa_\alpha A_\gamma \pa_\beta F_\mn - \theta^{\alpha\beta} \pa_\beta 
f^\alpha A_\alpha \pa_\gamma F_\mn
+ \cdots \ .
\end{multline} 
Correspondingly, the transformation of the right side of \refeq{e2.24} reads
\begin{multline}\label{e2.26b}
\bar\delta_f   (F_\mn + \theta^{\alpha\beta}  F_{\alpha\mu} F_{\beta\nu} -
\theta^{\alpha\beta} A_\alpha \pa_\beta F_\mn) =\\
 f^\gamma \pa_\gamma\hat F_\mn + \pam f^\gamma
\hat F_{\gamma\nu} + \pan f^\gamma \hat F_{\mu\gamma} 
+ f^\gamma \theta^{\alpha\beta} 
\pa_\alpha A_\gamma \pa_\beta F_\mn - \theta^{\alpha\beta} \pa_\beta 
f^\alpha A_\alpha \pa_\gamma F_\mn\\
+ \theta^{\alpha\beta} \pa_\alpha f^\gamma (F_{\gamma\mu} F_{\beta\nu} - 
F_{\gamma\nu} F_{\beta\mu})
+ \cdots \ .
\end{multline} 
\end{subequations}
The difference between the two appears in the last entry of \refeq{e2.26b}, which can be
rewritten as $\theta_f^{\gamma\beta} (F_{\gamma\mu} F_{\beta\nu} - 
F_{\gamma\nu} F_{\beta\mu})$. Since the parenthetical expression is antisymmetric in
$(\gamma,\beta)$, the above also equals $\delta_f \theta^{\gamma\beta} F_{\gamma\nu}
F_{\beta\mu}$ and vanishes for transformations that preserve~$\theta$. Thus the two forms
of  the gauge-covariant \ct s \refeq{e2.26} in fact coincide.

\section{Conclusion}
 When transformations of  noncommuting  coordinates are suitably restricted, we can
define gauge-covariant     rules for the transformation of gauge fields, just as in the
commuting case. To begin, the \ct\ is taken to be affine; then the gauge-covariant rule
\refeq{e2.15} for transforming the noncommuting vector potential/connection emerges with
the help of the Seiberg-Witten equation. When the \ct\ is further restricted so that it leaves
the noncommutativity tensor~$\theta$ invariant, the gauge-covariant transformation rule
extends to the field strength/curvature, \refeq{e2.19}. For these transformations, the
commutator algebra \refeq{e2.22} and  \refeq{e2.23} closes up to a \gt\ -- behavior familiar
from the commutative case\footnote{Of course with the further restriction that the
diffeomorphism leave $\theta$  unchanged, the covariant transformation law~\refeq{e2.15}
for the potential/connection is established directly from  \refeq{e2.4} -- the Seiberg-Witten
equation is not needed in
\refeq{e2.13} because $\delta_f \theta$ is taken to vanish.}. 

Our investigation concerns the kinematics of coordinate transformations. The dynamical
issue of identifying transformations that leave the equations of motion invariant and lead to
conserved quantities is addressed in a separate paper~\cite{8}. There it is established that in
Noether-theorem derivations of conserved charges, $\theta$ should not be transformed, in
concert with the position  taken in our paper. 

\subsection*{Acknowledgments}

R.J. has benefitted from conversations with H.~Grosse and V.P.~Nair. The research is
supported in part by funds provided by the U.S. Department of Energy (D.O.E.) under
cooperative research agreements
Nos.\ DE-FC02-94-ER40818 and  DE-FG02-91-ER40676.


\begin{thebibliography}{99}
\bibitem{6} The lack of complete closure in the commutator algebra prevents ``integrating''
the infinitesimal transformation to reach a finite transformation. A suggestion for
path-dependent finite transformations is given in~\protect\cite{1}.
\end{thebibliography}
\end{document}